\documentclass[9pt,twocolumn,twoside]{osajnl}

\journal{optica} 

\setboolean{shortarticle}{false}
\usepackage{caption}
\usepackage{lineno}
\usepackage{braket}
\usepackage{bm}
\usepackage{amsmath}
\usepackage{float}
\definecolor{R1}{rgb}{0,0.0,0.00}
\definecolor{R2}{rgb}{0.0,0.0,0}
\definecolor{R3}{rgb}{0.0,0,0}

\title{On-demand harnessing of photonic soliton molecules}

\author[1,2,*]{Shilong Liu}
\author[1,+]{Yudong Cui}
\author[2]{Ebrahim Karimi }
\author[3,4]{Boris A. Malomed}

\affil[1]{State Key Laboratory of Modern Optical Instrumentation, College
of Optical Science and Engineering, Zhejiang University,Hangzhou, Zhejiang
310027, China}
\affil[2]{Department of Physics, University of Ottawa, 25 Templeton Street, Ottawa, Ontario, K1N 6N5 Canada}
\affil[3]{Department of Physical Electronics, Faculty of
Engineering, and Center for Light-Matter Interaction, Tel Aviv University, Tel Aviv 69978, Israel}
\affil[4]{Instituto de Alta Investigaci\'{o}n, Universidad de Tarapac\'{a},
Casilla 7D, Arica, Chile}

\affil[*]{Corresponding author:  sliu7@uottawa.ca}
\affil[+]{Corresponding author: cuiyd@zju.edu.cn}

\dates{Compiled \today}
\doi{\url{http://dx.doi.org/10.1364/optica.XX.XXXXXX}}

\begin{abstract}
Soliton molecules (SMs) are fundamentally important modes in nonlinear optical systems. It is a challenge to experimentally produce SMs with a required temporal separation in mode-locked fiber lasers. Here, we propose and realize an experimental scenario for harnessing SM dynamics in a laser setup. In particular, we tailor SMs in a mode-locked laser controlled by second-order group-velocity dispersion and dispersion losses: the real part of dispersion maintains the balance between the dispersion and nonlinearity, while the dispersion loss determines the balance of gain and losses. The experimental
results demonstrate that the dispersion loss makes it possible to select desired values of the temporal separation (TS) in bound pairs of SMs in the system. Tunability of the SM's central wavelength and the corresponding hysteresis are addressed too. The demonstrated regime allows us to create multiple SMs with preselected values of the TS and central wavelength, which shows the potential of our setup for the design of optical data-processing schemes.

\end{abstract}
\setboolean{displaycopyright}{true}

\begin{document}

\maketitle

\section{Introduction}

Photonic \textquotedblleft soliton molecules" (SMs) are bound states of two
or more fundamental solitons, whose existence and stability are determined
by the separation and phase shift between the constituents ~\cite%
{mitschke1987experimental,malomed1991bound,akhmediev1997multisoliton,Tang2001}%
. For a stable SM, the relative phase of the bound solitons is fixed in the
course of the transmission. Nevertheless, the variation of the phase makes
it possible to observe complex dynamical effects, such as vibrations of the
\textquotedblleft molecule"~\cite{grapinet2006vibrating}, stepwise evolution
~\cite{herink2017real}, and phase sliding~\cite{krupa2017real,wang2019optical}%
. Direct studies of these effects have been recently made possible with the
help of the single-shot measurement technique, enabled by the time-stretch
dispersive Fourier transform (TSDFT)~\cite%
{berger2005complete,mahjoubfar2017time,peng2018real,liang2020real,chernysheva2020real}%
. Besides the relative phase, the temporal separation (TS) between the bound
solitons in the SM is another basic degree of freedom. It directly affects
the force of the interaction between the solitons, which is strong for TS
taking values comparable to the single-soliton's width, and weak for
essentially larger separations~\cite%
{malomed1991bound,Malomed1993,gui2018soliton,wang2019optical,he2019formation}%
. Much interest has been drawn to the TS degree of freedom due to the fact
that it affects numerous phenomena, including nonlinear evolution~\cite%
{krupa2017real,herink2017real,liu2018real}, short- and long-range
interactions~\cite{akhmediev1997multisoliton,he2019formation}, SM complexes%
~\cite{wang2019optical}, and the Hopf bifurcation which excites robust
intrinsic oscillations in SMs~\cite%
{sakaguchi2018stationary,grapinet2006vibrating}. The TS also plays the basic
role in the development of various SM-based optical applications, such as
\textquotedblleft multialphabetic" encoding for communications~\cite%
{stratmann2005experimental}, optical switching~\cite{kurtz2020resonant}, and
all-optical storage~\cite{pang2016all}. These studies make it necessary to
develop means for \textquotedblleft on-demand" manipulations of bound states
of solitons in nonlinear dissipative systems.
In particular, a machine-learning optimization algorithm, which helps to create a bound state of two solitons with a required  TS in the fiber laser, was reported very recently~\cite{girardot2022demand}.

Passively mode-locked fiber lasers offer a highly efficient platform for the
generation of the bound states. Many experiments performed in such systems
have produced stable SMs with the fixed relative phase~\cite%
{Tang2001,grelu2002phase,gui2018soliton}, and TS corresponding to tightly or
loosely bound \textquotedblleft molecules"~\cite%
{Tang2001,grelu2002phase,herink2017real,zhu2019observation,he2019formation}.
Such SMs were predicted by theoretical models based on the nonlinear Schr%
\"{o}dinger/Ginzburg-Landau equations~\cite{kivshar1989dynamics,
malomed1991bound,akhmediev1997multisoliton}. However, controllable
manipulations of SMs in a nonlinear dissipative system is a challenge to the
experiment. In particular, it is not easy to produce SMs with a
predetermined value of TS in a mode-locked fiber laser. Although several SMs
with different separations can be created by varying parameters of the fiber
cavity, \textit{viz}., the polarization or pump power, it is difficult to
exactly adjust the TS, or switch SMs between states with certain TSs
\textquotedblleft at will"~\cite%
{grelu2002phase,herink2017real,gui2018soliton,krupa2017real}. In particular,
the ability to produce bound states with predetermined (\textquotedblleft
on-demand") values of the separation is important for SM-based optical
data-processing schemes, communications~\cite{stratmann2005experimental},
optical switching~\cite{kurtz2020resonant}, storage~\cite%
{leo2010temporal,pang2016all}, and soliton trapping~\cite{jang2015temporal}.
Recently, the technique of active pump modulation was adopted to efficiently
manipulate SMs in mode-locked fiber lasers \cite{kurtz2020resonant,he2021synthesis}. In particular, all-optical switching of
two SMs was realized by dint of the frequency-swept pump modulation~\cite%
{kurtz2020resonant}. Actually, the pump modulation is a powerful technique
to control the SMs, while pump perturbations may lead to unwanted effects,
such as generation of higher-order harmonics~\cite{kurtz2020resonant}.

In this paper, we propose and experimentally realize a scenario allowing one
to perform on-demand manipulations of photonic SMs in a passively
mode-locked laser. Different from the external pump modulation, our method
relies on directly varying parameters of the nonlinear passive system,
\textit{viz}., the group-velocity dispersion (GVD)\ and dispersion losses,
to switch the SM from one state to another, with predetermined values of the
TS. The method works via controlling the complex dispersion: its real part
determines the balance between the GVD and nonlinearity in the mode-locked
laser, while the imaginary (dispersion-loss) part maintains the compensation
of the frequency-dependent losses by the gain. Studies of manipulations of
the dispersion had a long history in the course of the development of the
dispersion management for mode-locked lasers~\cite%
{malomed2006soliton,turitsyn2012dispersion,woodward2018dispersion,runge2020pure, iegorov2016direct,wei2020harnessing}%
.
\textcolor{R1}{In particular, a fourth-order soliton was recently created by
engineering both second- and fourth-order GVD in a mode-locked
laser system~\cite{runge2020pure}.  The result shows the imposed dispersion
in the cavity largely affects the nature of the soliton pulse, such as the
actual temporal profile shape. } However, the dispersion losses~\cite%
{akhmediev1997multisoliton,komarov2005multistability,sakaguchi2018stationary}
have drawn less attention in studies of nonlinear dissipative systems,
especially as concerns soliton pairs. Effects akin to the dispersion losses
were used for shaping temporal pulses in linear systems~\cite%
{monmayrant2010newcomer,weiner2011ultrafast}. Here, we demonstrate that
additional contributions to the dispersion losses play an important role in
the formation of the photonic SMs. In particular, in a specific range of
parameters, a large contribution to the imaginary part of the complex
dispersion increases the TS between the bound solitons. Experimentally, we
employ a $4f$ pulse shaper placed in the cavity which uses a hologram, to
tailor both GVD and the dispersion losses via encoding phase and amplitude
distributions in the hologram. The setup can realize producing multiple SMs
on-demand in the range of TS from $3.014$ to $5.478$ ps. By employing the
TSDFT technique, we observe the SM's spectral dynamics and demonstrate
the realization of quaternary encoding and optical switching between four SMs
with different temporal sizes. Further, by slightly moving the hologram in
the spatial light modulator, the tunability of the wavelength is realized,
and essential hysteretic phenomena are revealed.
\textcolor{R1}{Thus, the
hologram with the adjustable structure and position is the main tool that
is used in this work to realize efficient control of SMs in the fiber laser.}
The results not only offer a robust way to manipulate SMs in the mode-locked
fiber system, but also supports a new tunable degree of freedom for the
experimental control of nonlinear dissipative systems. The multiple SMs
reported in this work may find essential applications to the design of
\textquotedblleft multi-alphabetic" communications, optical switching, and
storage also in molecular spectroscopy.

\section{Methods for on-demand generation of SMs}

\begin{figure}[tbh]
\centering
\includegraphics[width=9cm]{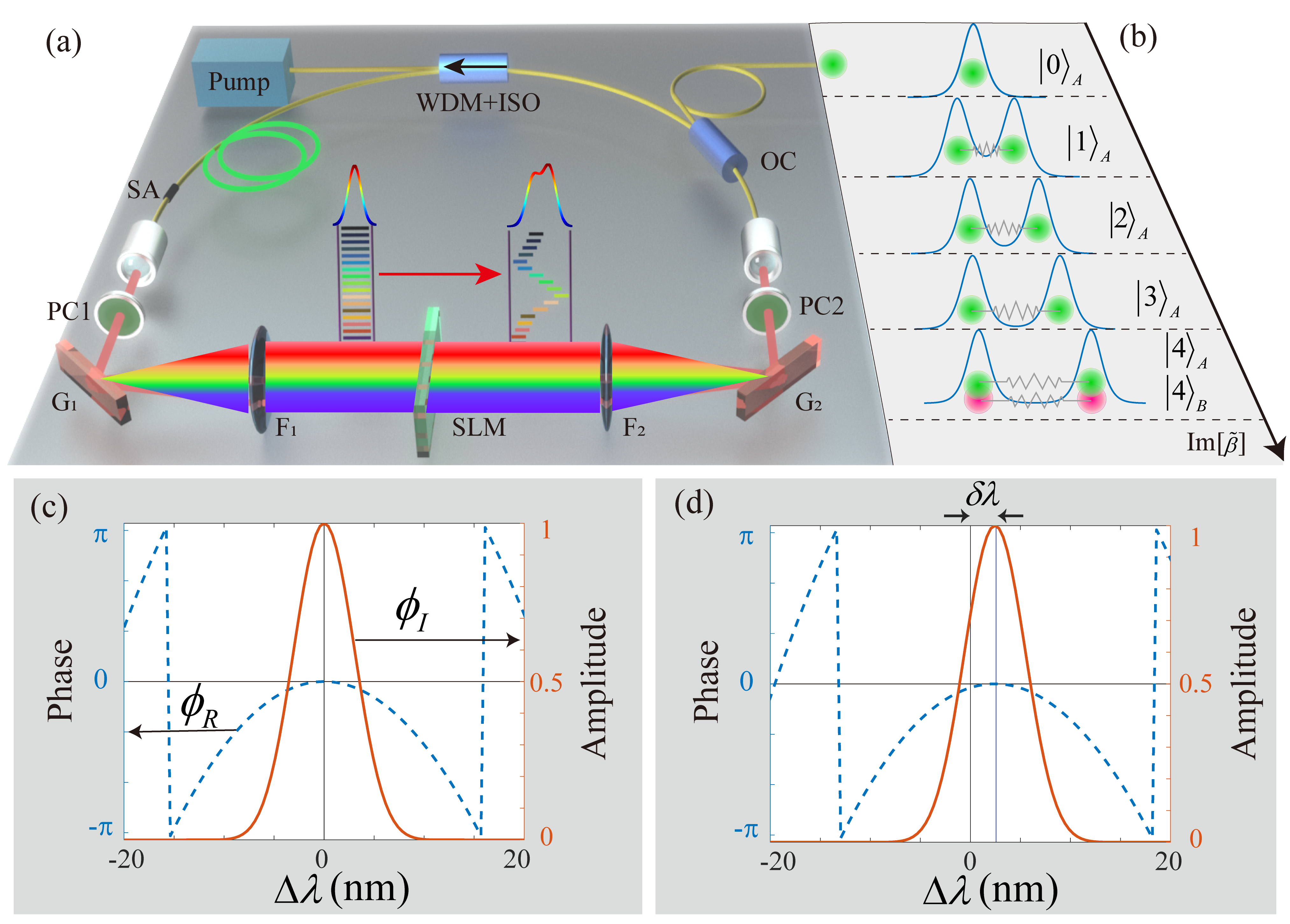}
\caption{The scheme for the generation of SMs (soliton molecules). (a): The
layout of the mode-locked fiber laser. Here, \textrm{Pump} stands for the
pump source at $980$ nm; WDM+ISO is the wavelength-division multiplexer for $%
980$ nm and $1560$ nm combined with the isolator for $1560$ nm; \textrm{SA}
is the saturable absorber; PC1(2) are polarization controllers
\textcolor{R2}{that includes one half-wave and two quarter-wave
plates};
and OC is the output coupler.
\textcolor{R1}{The
pulse
shaper
system
includes two
infrared
optical
gratings G$_{1,2}$
(the
grating
period is 940
gr
mm$^{-1}$),
input and
output lenses
$F_{1,2}$(the
focal
length is 125
mm), and
an
infrared
spatial light
modulator (SLM). It
has
the
resolution of
$1920 \times
1080$
and the
pixel pitch of 6.4 $\mu
m$.}
(b): The output SMs built of two solitons with the $\mathrm{sech}(t)$
profile, corresponding to different values of the spectral filtering, $%
\mathrm{Im}\left( \tilde{\protect\beta}_{2,\mathrm{holo}}\right) $. Here, $
\ket{m}_{A,B}$ are symbols for SMs with different values of TS (temporal
separation between the bound solitons), while subscripts $A$ and $B$
represent SMs with different central wavelengths. (c): Phase and amplitude
distributions of the encoding factor, given by Eq. (\protect\ref{E2}) with $%
\tilde{n}=10\left( 1+i\right) $ and $\protect\delta \protect\lambda =0$, in
the case when only the second-order GVD and dispersion losses are presented.
The distributions are displayed versus the offset of the operational
wavelength from the central point, $\Delta \protect\lambda =\protect\lambda %
-1554$ nm. (d): Same as in (c), but for the setup with the central
wavelength shifted by $\protect\delta \protect\lambda =2.5$ nm.}
\label{F1}
\end{figure}

Figure 1 illustrates the principle of manipulating the photonic SMs
determined by the effects of GVD and dispersive losses. Panel 1(a) shows the layout of the passively mode-locked laser, which includes fiber-guided and
free-space sections. The former one consists of a $3.3$ m long erbium-doped
fiber providing the necessary gain, a homemade saturable absorber providing
the mode-locking in the cavity, and the output coupler extracting $10\%$ of
the power from the fiber cavity. In the free-space section, we employ a $4f$
pulse shaper to morph the amplitude and phase of the spectrum. The total
length of the cavity is $\approx 19.88$ m, which includes $1.1$ m in the
free space for spectrum shaping. The measured cavity's fundamental
repetition rate is $10.458$ MHz. The net dispersion of the fiber cavity is $%
-0.2672$ ps$^{2}$, i.e., the mode-locked laser operates in the anomalous GVD
region.

The generation of SMs in the mode-locked fiber laser can be adequately
modeled by the complex Ginzburg-Landau equations (CGLEs)~\cite%
{malomed1991bound,akhmediev1997multisoliton,song2019recent}. The appropriate
CGLE for amplitude $A\left( z,t\right) $ of the output optical pulse is~\cite%
{skryabin2010colloquium,runge2020pure}
\begin{equation}
\frac{{\partial A}}{{\partial z}}=i\sum\limits_{k}{\frac{{{{\tilde{\beta}}%
_{k}}}}{{k!}}}{\left( {i\frac{\partial }{{\partial T}}}\right) ^{k}}A+\frac{{%
g-\alpha }}{2}A+i\gamma |A{|^{2}}A+\mathrm{H.O.E}.  \label{E1}
\end{equation}%
Here, $z$ and $t$ are the propagation distance and local time; $\gamma $ is
the Kerr parameter; $g$ and $\alpha $ represent the linear gain and loss,
respectively;
\textcolor{R3}{and  $\mathrm{H.O.E}$ stands for terms
representing higher-order nonlinear effects, which includes the saturation
of the nonlinear gain, saturation of the nonlinear refractive index, etc.
The higher-order terms are essential for the normal operation of the system}~
\cite{sakaguchi2018stationary,akhmediev1997multisoliton}.
\textcolor{R1}{In this model,
$\tilde{\beta}_{k}=\mathrm{Re}(\tilde{\beta}_{k})+i\,\mathrm{Im}(\tilde{\beta}_{k})$
includes GVD,  $\mathrm{Re}(\tilde{\beta}_{k})$, and the
dispersion losses, $\mathrm{Im}(\tilde{\beta}_{k})$,  of  orders $k=2,3,4,...$.
In this paper, $\tilde{\beta}_{k}$ are referred to as $k$-th order complex dispersion
coefficients.  In particular, $\tilde{\beta}_{2}$
represents the complex second-order dispersion. } Its GVD part, $\mathrm{Re}(%
\tilde{\beta}_{2})$, plays the central role in the operation  of the
dispersion management in mode-locked lasers~~\cite%
{malomed2006soliton,turitsyn2012dispersion,woodward2018dispersion,runge2020pure}%
. However, the dispersion-loss part, $\mathrm{Im}(\tilde{\beta}_{2})$, being
a necessary ingredient for the creation of two- and multi- soliton bound
states in dissipative nonlinear systems~\cite%
{malomed1991bound,akhmediev1997multisoliton,akhmediev1998stable,komarov2005multistability,sakaguchi2018stationary}%
, was less explored experimentally, especially in the context of SMs.
\textcolor{R3}{In the traditional mode-locked fiber laser system without the
addition filter, it is related to the gain bandwidth $\Omega _{g}$,
\textit{viz}., $\mathrm{Im}(\tilde{\beta}_{2})=g/2\Omega _{g}^{2}$
\cite{song2019recent}}. In our system, $\tilde{\beta}_{k}$ can be precisely
adjusted by means of the $4f$ pulse shaper, that transforms the temporal
pulse into the frequency domain, and then back into the temporal one. Using
the spatial light modulator (SLM) installed at the center of the pulse
shaper, it is possible to perform flexible management of the complex
dispersion coefficient of an arbitrary order (details of the pulse shaper
could be found in \textbf{Section 1 of the Supplementary Material}). In this
paper, the second-order term is addressed in the main text, the higher-order
terms being considered in\textbf{\ Section A of the Appendix and Section 2
of the Supplementary Material}. As mentioned above, $\mathrm{Re}\left( {%
\tilde{\beta}_{2}}\right) $ provides the dispersion which balances the
nonlinearity in the setup of the mode-locked laser, which is the ordinary
condition necessary for the formation of optical solitons \cite%
{agrawal2000nonlinear,kivshar2003optical,runge2020pure}, while the loss
part, $\mathrm{Im}\left( {\tilde{\beta}_{2}}\right) $, is responsible for
maintaining the balance of the gain and losses in the system \cite%
{grelu2012dissipative}. To distinguish the complex dispersion coefficients
in different sections of the system, we define $\tilde{\beta}_{2,\mathrm{%
fiber}}$ as the coefficient in the fiber section, and $\tilde{\beta}_{2,%
\mathrm{holo}}$ as one imprinted by the hologram in the $4f$ pulse shaper.

Our experimental findings and simulations \textbf{(see Section A in Appendix)%
} demonstrate that the dynamics of the SMs can be tailored by adjusting the
real and imaginary parts of ${\tilde{\beta}_{2,\mathrm{holo}}}$. In this
context, Fig. 1(b) shows that TS between the bound solitons gradually
increases, following the increase of $\mathrm{Im}{(\tilde{\beta}_{2,\mathrm{%
holo}})}$ in a certain range. In particular, only the conventional single
soliton, symbolized by $\ket{0}$, is possible at $\mathrm{Im}(\tilde{\beta}%
_{2,\mathrm{holo}})=0$. With the increase of $\mathrm{Im}(\tilde{\beta}_{2,%
\mathrm{holo}})$, the system can create SMs on-demand with different values
of TS, $\ket{m}(m=1,2,3...)$ (see the illustration in Fig. 1(b)). In
addition, the setup could change the color (central wavelength) of SMs,
while keeping the TS constant. The latter option is shown in the bottom line
of Fig. 1(b) by means of labels $\ket{4}_{A,B}$, where $A$ and $B$ represent
the central wavelength of two SMs. Experimentally, both the GVD and
dispersion-loss constituents of $\tilde{\beta}_{2,\mathrm{holo}}$ can be
manipulated through the phase and amplitude distribution of the hologram in
the SLM. The hologram's characteristic is defined as $H(x)=\phi _{R}(\mathrm{%
Re}(\tilde{\beta}_{2,\mathrm{holo}}),x)\cdot \phi _{I}(\mathrm{Im}(\tilde{%
\beta}_{2,\mathrm{holo}}),x)\cdot \phi _{G}(d,x)$, which includes three
factors used for the implementation of the precise manipulations. Here, $%
\phi _{R}(\mathrm{Re}(\tilde{\beta}_{2,\mathrm{holo}}),x)$ and $\phi _{I}(%
\mathrm{Im}(\tilde{\beta}_{2,\mathrm{holo}}),x)$ represent real and
imaginary terms (corresponding to the phase and amplitude encoding,
respectively), while $\phi _{G}(d,x)$ represents an optical grating that
helps to avoid crossover with the zero-order diffraction \cite%
{bolduc2013exact,liu2019classical}. For the description of the experiment,
we define the dimensionless complex coefficient $\tilde{n}$ to encode the
hologram by means of its real and imaginary parts:
\begin{subequations}
\begin{align}
\mathrm{Re}\left( {\tilde{\beta}_{2,\mathrm{holo}}}\right) & ={k_{R}}\cdot
\mathrm{Re}\left( {\tilde{n}}\right)  \label{Za} \\
\mathrm{Im}\left( {\tilde{\beta}_{2,\mathrm{holo}}}\right) & ={k_{I}}\cdot
\mathrm{Im}\left( {\tilde{n}}\right) .  \label{Zb}
\end{align}%
Here, $k_{R}=-0.2170\times 10^{-3}$ ps$^{2}/$m and $k_{I}=0.9309\times
10^{-3}~$ps$^{2}/$m are two appropriate constants that are selected
empirically in accordance with the experiment;
\textcolor{R2}{the values are subject to the
constraint imposed by the capacity of the pulse shaper
(see the details in the \textbf{Section 1 of the Supplementary
Material})}. $\mathrm{Re}\left( {\tilde{n}}\right) $ represents the positive
or negative number associated with GVD, while $\mathrm{Im}\left( {\tilde{n}}\right) $ connects to
the dispersion losses, hence it may be only positive. Following these
definitions and ignoring the higher-order dispersion (which is addressed in
\textbf{Section 2 of the Supplementary Material}), the encoding part of the
hologram is written as
\end{subequations}
\begin{equation}
{\phi }\left( {{{\tilde{\beta}}_{2,\mathrm{holo}}}}\right) \simeq \exp \left[
{i{{\tilde{\beta}}_{2,\mathrm{holo}}}(\tilde{n}){{L_{\mathrm{cavity}}(\omega
-\delta \omega )}^{2}}/2}\right] ,  \label{E2}
\end{equation}%
where $L{{{_{\mathrm{cavity}}}}}$ is the total length of the cavity (as
mentioned above, it is $19.88$ m in our experimental setup), $\omega $
represents the optical frequency corresponding to the wavelength at the
spatial position of the SLM, and $\delta \omega $ is an offset of the
central frequency, which leads to a shift for the effective gain wavelength
range of the mode-locked fiber laser. The offset, which can be implemented
experimentally by slightly displacing the position of the hologram, provides
the precise tunability of the central wavelength for the output SMs.

Figure 1(c) displays an example of the distribution of the phase (blue) and
amplitude (red) produced by Eq. (\ref{E2}) with $\delta \omega =0$, where $%
\tilde{n}=10\left( 1+i\right) $ is set. Such one-dimensional curves can be
expanded for the two-dimensional hologram by adding the optical grating (see
\textbf{Section 2 in Supplementary Material} for details in hologram).
Figure 1(d) shows the distribution produced by Eq. (\ref{E2}) with $\delta
\omega $ corresponding to the wavelength shift $\delta \lambda =2.5$ nm.

In the present regime, the real and imaginary parts of $\tilde{\beta}_{2,%
\mathrm{holo}}$ can be independently determined by the amplitude and phase
structure of the hologram. The ability to choose these parameters offers a
robust scheme for manipulating SMs in the mode-locked system. In turn, the
output produced by the system, in the form of the photonic SMs, offers the
platform for encoding data in the optical domain, including options for
multiple encoding and optical switching~\cite%
{stratmann2005experimental,kurtz2020resonant}. Below, we report experimental
results for manipulations of TS in the bound soliton pairs, quaternary
encoding, and switching between several SMs, as well as the wavelength
tunability of SMs and a hysteresis effect for them.

\section{Results}

\subsection{Manipulations of the TS (temporal separation) for SMs (soliton
molecules) in the mode-locked laser}

\begin{figure}[!htb]
\centering
\includegraphics[width=8cm]{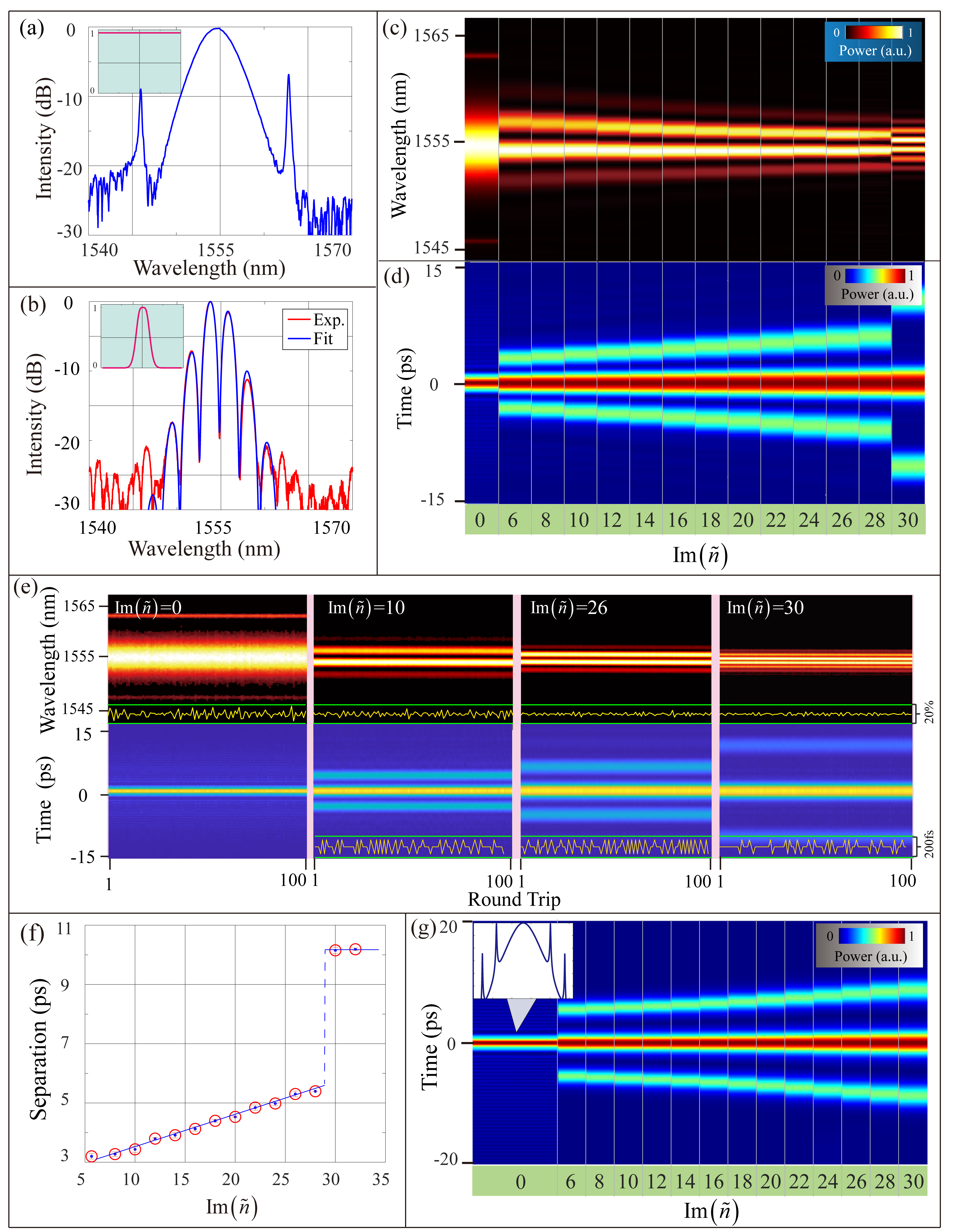}
\caption{The generation of the SMs (soliton molecules) in the experiment.
(a) and (b): The spectral intensity in the single-soliton ($\tilde{n}=0$)
and SM ($\tilde{n}=10\left( 1+i\right) $) regimes, respectively. \textcolor{R1}{The inserts in the top left corner are the applied amplitude profile of hologram in SLM, respectively. }
(c) and (d): Spectral intensities measured by the optical spectrum analyzer and
their Fourier transform, which is also the autocorrelation function,
obtained for \textrm{Im}$\left( \tilde{n}\right) $ varying from $0$ to $30$.
(e): The evolution of spectral intensities measured by TSDFT, where the
recording length is $100$ round trips; the corresponding autocorrelation
functions are shown at the bottom.
\textcolor{R1}{Here, the insets at the bottom of the panels show
oscillations of differences in the energy and separations (yellow
lines), and green horizontal lines represent limit values. The basic
separation is $\pm$ 100 fs.} (f): The TS\ (temporal separation between the
bound solitons) extracted from the spectrum recorded by TSDFT.
\textcolor{R2}{Here, we have added the triple standard deviation  to each data point.
For instance, it is $66$ fs for $\tilde n=10(1+i)$. Circles
surrounding the points are added for better
visualization.} (g): Results produced by
simulations of the theoretical model based on the CGLE and pulse-shaping
equations (see \textbf{Section A in Appendix}). The inset in the top left corner is the spectral intensity produced by the simulations in the single-soliton regime, where the bandwidth is $4.37$ nm.}
\label{F2}
\end{figure}

We first adjust the mode-locked fiber laser to work in the regime of
producing individual solitons. To this end, first, the hologram parameter is
taken as $\tilde{n}=0$, which represents a simple transmitting mirror. \textcolor{R2}{Next,
we optimize parameters in the cavity, \textit{viz}., the pump power and
polarization, to provide for the output in the form of a single soliton.} The output spectrum of the single-soliton is shown in Fig. 2(a), where the bandwidth is $4.40$ nm. Two strong spectral sidebands observed in the spectrum are the usual feature of the conventional single-soliton generation regime~ \cite{malomed1987,kelly1992characteristic,liu2018real}. Using a
homemade autocorrelation-FROG system, we reconstruct the solitons' temporal
and spectral profiles and their envelope phase (see \textbf{Section 3 in
Supplementary Material}). The so reconstructed pulse width is $670$ fs, and
its chirp~\cite{agrawal2000nonlinear} is very low.

Next, we switch the hologram, changing the complex control parameter, $%
\tilde{n}$, defined above in Eqs. (\ref{Za}) and (\ref{Zb}), so as to
provide the generation of soliton pairs. The SMs could be observed from
spectral fringes, that, in turn, are produced by the optical spectrum
analyzer~\cite{berger2005complete,herink2017real}. We set $\mathrm{Re}(\tilde{%
n})=10$ and increase $\mathrm{Im}(\tilde{n})$, starting from $0$. The output
spectra are unstable in the interval from $0$ to $6$, which means many
possible different spectra
\textcolor{R2}{(an example is presented in
\textbf{Video 1 of Supporting Material})}. A stable SM appears when $\mathrm{%
Im}(\tilde{n})=6$. Later, the output features SMs, whose TS increases
linearly up to $\mathrm{Im}(\tilde{n})\approx 29$. Fig. 2(b) shows a power
spectrum representing the SM for $\tilde{n}=10\left( 1+i\right) $, where the red and blue lines represent, respectively, the experimental result and fitting
to it (for the fitting method, see \textbf{Section 4 in Supplementary Material}). Figure 2(c) shows the spectral distribution for different values
of $\mathrm{Im}(\tilde{n})$, \emph{viz}., $0$ and $6\leq \mathrm{Im}(\tilde{n%
})\leq 30$ with interval $\Delta \left( \mathrm{Im}(\tilde{n})\right) =2$.
The color-coding in this figure represents the relative spectral intensity,
which is normalized to take values from $0$ to $1$.

The spectral distributions provide the basic information for the SMs in the
temporal domain. In particular, the TS of the bound solitons is inversely
proportional to the spectral fringe period, and their relative phase can be
inferred from the envelope of the spectral interferogram \cite%
{herink2017real}. To directly find the TS, it is instructive to perform the
Fourier transform of the spectral intensity, which also represents the
autocorrelation function~\cite{berger2005complete,herink2017real}. Figure
2(d) exhibits the Fourier transform of the corresponding autocorrelation
function. The results show that the TS grows linearly with an increase of $%
\mathrm{Im}(\tilde{n})$ from $6$ to $28$. With the further increase of $%
\mathrm{Im}(\tilde{n})$, the SM becomes unstable, switching to another SM
state with a large TS, as observed in Fig. 2(d) for $\mathrm{Im}(\tilde{n}%
)=30$. This phenomenology is similar to that demonstrating the Hopf-type
bifurcation in two-soliton bound states in other systems \cite%
{ott2002chaos,grapinet2006vibrating,sakaguchi2018stationary}.  \textcolor{R2}{From the spectral interferogram of SMs, one can find that the dispersion loss strongly affects the separations, while it is practically not related to the relative phase between two soliton pairs. For the relative
phase, many previous works show that it may be affected by the pump power or PC orientation in the
mode-locked fiber system~\cite{herink2017real}.}
The full dynamical behavior of SMs supported by different holograms can be seen
\textbf{Video 1 in Supporting Material}.

To explore the stability of the SMs, we measured the single-shot spectrum by
means of the TSDFT  technique, which may also be used to perform real-time
multiple SMs encoding and switching
\textcolor{R1}{(the traditional
optical spectrum analyzer is unsuitable in this case, due to the slow scanning speed).}
For this purpose, the output optical solitons are injected into a dispersion
fiber ($5000$ m long) \cite{mahjoubfar2017time,cui2021xpm}, in which the
spectral profile is mapped into the time domain \cite%
{mahjoubfar2017time,torres2011space} (Also see \textbf{the Section 1 of the
Supplementary Material}). The recorded dispersive Fourier-transform spectra
are shown in Fig. 2(e), where we present the conventional single soliton (at
$\mathrm{Im}(\tilde{n})=0$) and three SMs (at $\mathrm{Im}(\tilde{n}%
)=10,26,30$) with different STs. In these measurements, the recording length
(shown on the horizontal axis) is up to $100$ round trips. Four subplots
displayed in the figure are the corresponding autocorrelation functions.
Figure 2(f) shows the average TS extracted from the dispersive
Fourier-transform spectra, where the blue line is the fitting result,
produced by the linear model in the following form: TS(ps) $=0.112\times $ $%
\mathrm{Im}{(\tilde{n})}+2.342$, $6\leq \mathrm{Im}(\tilde{n})\leq 28$.
Close to $\mathrm{Im}(\tilde{n})=29$, a large jump in the TS appears, and
then it does not change conspicuously with the further increase of $\mathrm{%
Im}(\tilde{n})$. Here, only the range of $\mathrm{Im}(\tilde{n})$ between $6$
and $28$ is considered, where the TS can be tailored at will from $3.014$ to
$5.478$ ps, and the SMs can be switched gently without a hysteresis. Such
features are essential and useful to perform quaternary encoding and optical
switching in the SMs, as shown in the next section.

\textcolor{R1}{ One can analyze the stability of the SMs from the single-shot spectrum
recorded by TSDFT. There are several methods to character the stability of the SM states.
In particular, the stability can be explored in the plane of the separation and relative phase \cite{afanasjev1997stability,akhmediev1997multisoliton,herink2017real}, through measuring the radio-frequency (RF) spectra~\cite{liu2021resolving,xian2020subharmonic}, or directly
observing variations of the energy and separation of the bound pulses~\cite{akhmediev1997multisoliton,liu2016discrete,peng2019breathing,liu2020visualizing,wang2020recent}.  By considering our demonstrations,  we show the variations in pulse energy and separations from the single-shot spectrum, which could directly reflect the stability of SMs.  For single-shot spectrum, we define the energy difference as $dQ_m/dm$, where ${Q_m} = \frac{1}{{{Q_{ave}}}}\int {{{\left| {A(m,\omega )} \right|}^2}d\omega }$ is the spectrum energy for  $m$-th round trip from TSDFT, and ${Q_{ave}} = \sum\limits_{m = 1}^M {\frac{1}{M}\int {{{\left| {A(m,\omega )} \right|}^2}d\omega } }$ is the average value among all the recording $M$ round trips. Therefore, the difference could be simply given as  $Q_{m}-Q_{m-1}$. Actually, the definition is the stable criterion for SMs in energy aspect~\cite{akhmediev1997multisoliton}.  We calculated the energy difference for our recording single-shot spectrum of $\mathrm{Im}(\tilde n)$ =0, 10, 26, and 30, which are shown at the bottom of Fig.  \ref{F2} (e). Here, two green lines set the basic energy level, which is $\pm$10\% for $Q_{ave}$; the yellow lines give the oscillations of the energy difference.  The similar calculation is performed to the separations that simply defines $\tau_{m}-\tau_{m-1}$ . We add the separation difference at the below of autocorrelation functions. Here, two green basic lines set $\pm$100 fs, which correspond to the same level of the temporal resolutions from the TSDFT system (see the \textbf{Section 1 of the Supplementary Material} ). The yellow line represents the extracting difference, which is slightly below the set level $\pm$100 fs. These results illustrate the SMs ($\mathrm{Im}(\tilde n)=0,10,26,30$) are stable at the current recording range (100 round trips). It is enough for the demonstrations because the multiple SMs based encoding regime only accounts for 20 round trips.  By the way, in the \textbf{Section 1 of the Supplementary Material} , we add one radio-frequency (RF) spectra for SM state of $\tilde n=10(1+i)$. It shows the average signal-noise level could be up to around 50 dB.}

\begin{figure}[tb]
\centering
\includegraphics[width=8.2cm]{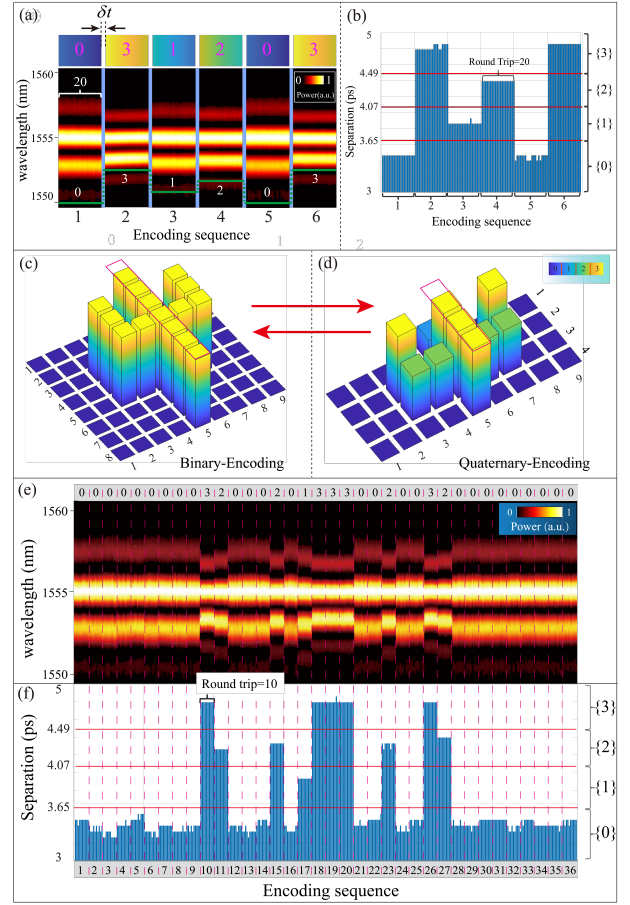}
\caption{The SM-based quaternary encoding and switching. (a) The measured
single-shot spectrum for string $\left\{ 0|3|1|2|0|3\right\} $, produced by the TSDFT technique with $20$ round trips. (b): The corresponding TSs of six
SM states. (c)-(d): A \textup{$\protect\psi $}-shaped pattern in the binary
and quaternary encoding systems, respectively. (e): The measured single-shot
spectrum for the \textup{$\protect\psi $}-shaped pattern with 36 encoding
holograms in the string of $\left\{
0000|0000|0320|0020|1333|0020|0320|0000|0000\right\} $, where each encoding
sequence is produced by TSDFT with $10$ round trips. (f): The corresponding
TSs extracted from the encoding sequence with 36 holograms.}
\label{F3}
\end{figure}

We have performed simulations of the dissipative nonlinear system shown in
Fig. \ref{F1}(a). For this purpose, we split the cavity into two main parts: in the fiber section, it is modeled by CGLE taken as Eq. \ref{E1}, while in the free-space section the model is provided by the pulse-shaping equation. Also, we need to address equations for the saturable absorber and gain from the erbium-doped fiber. Details are given in \textbf{Section A of Appendix}.
Parameters were first adjusted to those corresponding to the single-soliton regime in Fig. \ref{F2}(a). The result obtained by the simulations for the single-soliton spectrum is shown in the inset to Fig. \ref{F2}(g), where two
sidebands obviously belong to the spectral domain. The respective bandwidth is $4.37$ nm, which is in good agreement with the experimental observations ($4.40$ nm in Fig. \ref{F2}(a)). Next, we set $\tilde{n}=10+6i$, which leads to the
creation of bound soliton pairs. A similar simulation, for $\tilde{n}%
=10\left( 1+i\right) $, is presented in Fig. S1(b) of \textbf{Section A in
Appendix}. Fig. \ref{F2}(g) shows the autocorrelation function of the output produced by the simulations after 2000 round trips (the corresponding spectra are displayed in Fig. S1(c) of \textbf{Section A in Appendix}). From these results, it is seen that the TS of the bound states increases near linearly with the growth of $\mathrm{Im}(\tilde{n})$ in the range of $%
\mathrm{Im}({\tilde{n}})$ from $6$ to $28$, similar to what is observed in
the experiments, cf. Fig.  \ref{F2}(d). There are two conspicuous differences between simulations (Fig. \ref{F2}(g)) and experiments (Fig.  \ref{F2}(d)). One is the variation for the separations: the experiments show a linear increase of TS, while the simulations produce a nonlinear change. The other is the values of TS: the predicted and experimentally observed separations are not exactly equal for each Im$(\tilde n)$.
\textcolor{R1}{These differences between the
experiments and simulations are due to minor differences in the GVD and dispersion loss. They mainly originate from imperfections of the $4f$ pulse shaper, as concerns the coupling efficiency, encoding formats, and imaging precision (\textbf{Section 2 in Supplementary Material} addresses consequences of the imperfect encoding for the hologram).  The imperfections could lead to the appearance of higher-order dispersion and dispersion losses in the experimental setup.  If the higher-order term is added to the theoretical model, the difference between the experimentally observed and numerically predicted values of the TS becomes smaller.
For example, recent theoretical results have demonstrated that the third-order
dispersion strongly affects the separations of
SMs~\cite{sakaguchi2018stationary}. }
\textcolor{R3}{Further, the imperfect $4f$ pulse shaper can also limit the tuning range of
separations of SMs in the present system. Previous studies have also found that the linear tuning range
was limited -- for example, in the presence of the Hopf bifurcation~\cite{grapinet2006vibrating,sakaguchi2018stationary}.
If, as a result, the separation undergoes a jump with the variation of
the system's parameters, this implies instability or hysteresis of the SM.}  In additional simulations, which include the TOD term (not shown here in detail), we found that coexisting bound states appear,  especially for the larger value of Im$(\tilde n$), \textit{viz}.,  Im$(\tilde n)>35$.  This fact may quantitatively explain the jump that happens at larger Im$(\tilde n)$ . Generally, the theoretical model proposed here is not comprehensive for the comparison to the experiment. Indeed, not all experimental parameters could be exactly determined, and some assumptions are adopted to run the simulations. Nevertheless, the main features of the experimental results and their dependence on parameters are correctly predicted by the present model.

\subsection{SM-based quaternary encoding and optical switching}

Multiple encoding scenarios based on the use of SMs had been proposed long
ago, in the context of the multiple-soliton communications~\cite%
{stratmann2005experimental}, storage \cite{pang2016all}, soliton trapping~\cite%
{jang2015temporal}, and optical switching~ \cite{kurtz2020resonant}. Multiple
encoding helps to increase the information capacity in these applications.
\textcolor{R1}{However, there were few experiments to demonstrate multiple encoding. The main
issue is the difficulty in the preparation of an ``on demand" SM source.
Recently, the creation of two deterministic SMs was demonstrated with the help of the
pump-modulating technique~ \cite{kurtz2020resonant}. A challenging problem is to
efficiently manipulate more SMs and demonstrate possible regimes of multiple
encoding and switching.} The SM regimes revealed in this work offer new
possibilities to tailor the bound states of optical solitons, which can be
directly applied to the design of SM-based quaternary encoding and
optical-switching schemes~ \cite{stratmann2005experimental,kurtz2020resonant}.

Multiple SMs with different values of TS offer a possibility for multiple
encoding using different states $\ket{m}$, as shown in Fig. 1(b). Indeed, it
is seen from Figs. 2(b) and (d) that the system can produce many SM
varieties. For example, the TSs in the range of $3.014$ to $5.478$ ps
correspond to $\mathrm{Im}\left( \tilde{n}\right) $ from $6$ to $28$. To
avoid the crosstalk induced by the interaction between adjacent SMs \cite%
{tang2005soliton} and the TS jitter during the hologram switching, we select
four SMs, with $\mathrm{Im}\left( \tilde{n}\right) =10,14,18$, and $22 $, as
states $\{\ket{0}$, $\ket{1}$, $\ket{2}$, $\ket{3}\}$, respectively, for the
realization of quaternary encoding. Experimentally, these four states can be
controlled by appropriately using four holograms in the SLM; thus, the
optical switching can be performed by replacing the corresponding hologram.
In the experiment, we use four SMs to create a quaternary encoding scheme
that forms four optical bits. Here, we do not use the single-soliton regime (%
$\mathrm{Im}(\tilde{n})=0$) and SMs with a very large separation ($\mathrm{Im%
}(\tilde{n})\geq 30$). The main reason for the exclusion of such states is
the presence of strong hysteresis between them, which would disrupt the
implementation of the multiple-encoding and optical switching. Similar
hysteretic phenomena readily appear in the work with polarization- or
power-induced SMs~\cite{grapinet2006vibrating}. However, our experimental
results reveal no evident hysteresis in the above-mentioned range of $%
\mathrm{Im}\left( \tilde{n}\right) $ between $6$ and $28$.

Thus, we start by using the four above-mentioned SMs, $\ket{m}$ with $%
m=0,1,2,3$, and construct a simple test string $\left\{ 0|3|1|2|0|3\right\} $%
. The respective set of six holograms are consecutively displayed on the
SLM, switching them automatically in sequence with the time interval of $%
\delta t=$ $511$ ms, limited by the low operating speed of the
liquid-crystal SLM. Simultaneously, TSDFT records the corresponding
single-shot spectrum. In particular, the oscilloscope records and stores
four single-shot spectra for each hologram, in which the sampling number is
limited by the storage depth of the oscilloscope. Fig. \ref{F3}(a) shows an example of
the spectrum, with $20$ round trips recorded for each hologram. Further,
Fig. \ref{F3}(b) shows the corresponding values of TS in the bound state of two
solitons, as extracted from the single-shot spectrum. \textcolor{R2}{We conclude that the
single-shot spectrum is nearly stable in the framework of each encoding
sequence, and the corresponding TSs feature an acceptable jitter.}

To perform quaternary encoding by four states $\ket{m}$ with different
values $\tau $ of the TS, a robust criterion should be adopted to
distinguish values of $\tau $. Here, we sort them into four states: $\ket{0}$
($\tau \in \lbrack 3.0,3.64]$), $\ket{1}$ ($\tau \in \lbrack 3.65,4.06]$), $%
\ket{2}$ ($\tau \in \lbrack 4.07,4.48]$), and $\ket{3}$ ($\tau \in \lbrack
4.49,5.00]$), respectively. Labels attached to the right vertical axis in
Fig. \ref{F3}(b) shows the respective numbers of the so-coded states. According to
the TS values in Fig. \ref{F3}(b), the output state is indeed $\left\{
0|3|1|2|0|3\right\} $, being tantamount to the encoding sequence which the
procedure aimed to create.

We then consider a more complex encoding, \textit{viz}., the \textup{$\psi $}%
-shaped binary graph is shown in Fig. \ref{F3}(c), which includes $72$ digits in the
binary format ($00000000|00000000|00111000|00001000|01111111|00001000$ $%
|00111000|00000000|00000000$) However, using the SM encoding, it may be
reduced to a set of $36$ digits in the quaternary-encoding basis, as shown
in Fig. \ref{F3}(d), namely, $\left\{
0000|0000|0320|0020|1333|0020|0320|0000|0000\right\} $. We generate the
corresponding set of $36$ holograms and switch them automatically in
a sequence similar to how this was done above for the simple string, $\left\{
0|3|1|2|0|3\right\} $. Fig. \ref{F3}(e) shows the single-shot spectrum recorded
with the help of TSDFT, using $10$ round trips for each state sequence. The
criterion to distinguish the states $\ket{m}$ with $m=0,1,2,3$ is the same
as adopted in Fig. \ref{F3}(b). Analyzing the autocorrelation function, we could
identify the corresponding values of the separation of SMs, which is
presented in Fig. \ref{F3}(f), with the output pattern, $\left\{
0000|0000|0320|0020|1333|0020|0320|0000|0000\right\} $, being exactly the
same as the encoding sequence defining the \textup{$\psi $}-shaped target.
The dynamic switching results for the \textup{$\psi $}-shaped patterns in
the quaternary-encoding regime is presented in \textbf{Supporting Material
for Video 2}.

\begin{figure}[tb]
\centering
\includegraphics[width=8cm]{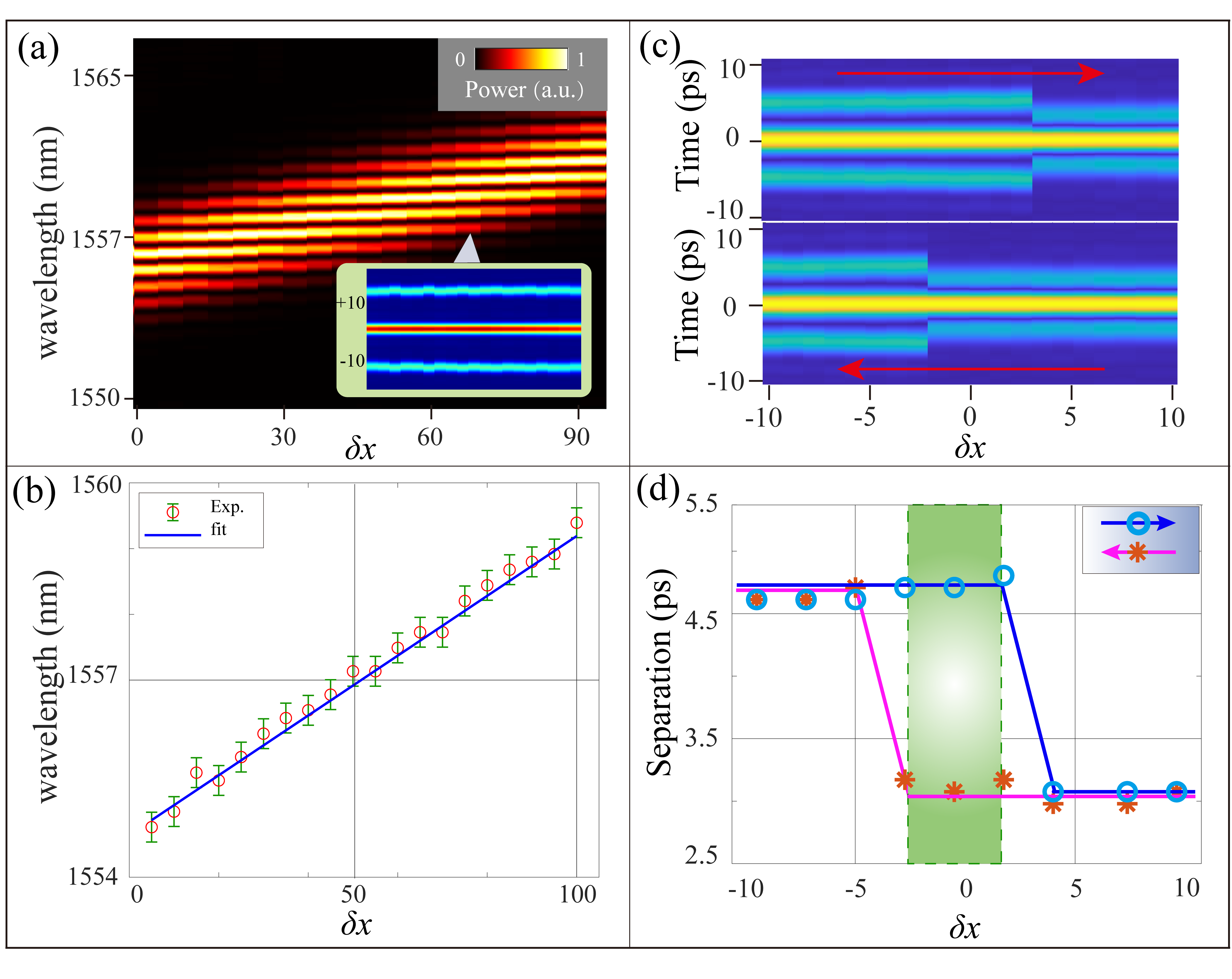}
\caption{The wavelength tunability (for $\tilde{n}=10+30i$) and hysteresis
(for $\tilde{n}=10+19i$) of the SMs. (a):
\textcolor{R2}{The
spectral
fringe
for SMs with different central
wavelengths, while
their
TS
(temporal
separation) is kept nearly
constant, with the
avergae
value
$\tau =12.544$
ps, and  average
variation $0.288$ ps.} The corresponding autocorrelation function is shown
in the right bottom corner. (b): The central wavelength $\protect\lambda _{0}
$, extracted from the experimental data, versus the relative position $%
\protect\delta x$ of the hologram. Here, we have added the measurement error $\pm 0.1134$ nm to each data point, which is the wavelength resolution  at $\sim$ 1555 nm of the employed optical spectrum analyzer. Panels (c) and (d) display a
wavelength-controlled hysteresis for the SMs, in terms of the TS. (c): The
autocorrelation function of the SMs, where the top and bottom plots
correspond to varying relative $\protect\delta x$ from left (shorter
wavelength) to right (longer wavelength), or from right (longer wavelength)
to left (shorter wavelength), respectively. (d) The TS data extracted from
(c). In these results, $\protect\delta x=1$ means that the actual spatial
shift of the hologram in SLM is 10 $\mu $m.}
\label{F4}
\end{figure}

The demonstrated SMs can realize multiple encoding based on the set of four
states in the quarternary system and switch between them with high
fidelity. Compared to the binary encoding system with $2^{n}$ options, the
quaternary one offers $4^{n}$, which is beneficial to perform dense coding
in optical communications, imaging, and computations~\cite%
{barreiro2008beating,corcoran2020ultra}. In our system, if one suitably
reduces the criterion interval distinguishing different bound soliton pairs,
the system could realize more SMs ($d$), which means it's possible to
perform higher-dimensional ($d^{n}$) encoding and switching.
\textcolor{R2}{In this case, a more  accurate technique should be employed to
characterize the separations of SMs, such as the balanced optical cross-correlation method~
\cite{song2020attosecond}, to show their time jitter on the fs scale.}

\subsection{The tunability of the central wavelength of the SMs and the
hysteresis}

\textcolor{R1}{The ability to tune the wavelength is needed in many applications, including optical sensing, metrology, and spectroscopy~\cite{wei2020harnessing,keller2003recent}.  The usual tuning methods employ the nonlinear wave
mixing~\cite{boyd2020nonlinear}, or the use of high-speed electric optical modulators. In a mode-locked fiber laser system, one usually changes the soliton's wavelength by adjusting the intra-cavity tunable filter~\cite{sun2010stable,xiang2018observation}. For SMs, there was no report of
high-precious tuning their wavelengths. The present setup may offer a simple platform to adjust the wavelength of SMs.}
According to Eq.~(\ref{E2}), $\delta \omega =\left( 2\pi c/\lambda
_{0}^{2} \right)\delta \lambda  $ represents the shift with respect to the
central wavelength $\lambda _{0}$, which leads to a shift of the effective
gain area in the cavity, and can be used to realize the precise tunability.
In the experiment, by shifting the hologram in the $x$ direction by $\delta
x $, which affects the real and imaginary parts of GVD in SLM through the
space-frequency relation of the $4f$ shaping system \cite%
{monmayrant2010newcomer}, the effective gain area displaced by $\delta
\omega $ will drag the central wavelength of the intracavity pulse (see
\textbf{Fig. \ref{F1}(d)}).

Figure \ref{F4} displays the experimental results concerning the shift of the
central wavelength of the SMs. We first lock the fiber laser to a stable SM
operation corresponding to $\tilde{n}=10+30i$, for which the central
wavelength is $1554.6$ nm. Then, after moving the position of the hologram
in SLM by $\delta x$, the corresponding normalized spectrum is recorded by
the optical spectral analyzer, as shown in Fig.  \ref{F4}(a). We further apply the
Fourier transform to the spectrum, to produce the autocorrelation functions,
which are shown in the inset to Fig.  \ref{F4}(a). These results demonstrate that the SM's central wavelength changes linearly, while the period of the spectral fringes remains nearly constant,
\textcolor{R2}{corresponding to the TS which takes values
$12.544$  $\pm$0.288 ps}. Figure \ref{F4}(b) shows the relation between $\delta
x$ and the wavelength shift extracted from Fig. \ref{F4}(a), where red circles represent data of experimental measurements, and the blue solid line is the fitting produced by the linear model, namely, $\lambda _{0}({\mu m}%
)=0.0455\times \delta x+1.5546$. Thus, our results exhibit the wavelength shift by $4$ nm (from $1555$ nm to $1559$ nm), with the linear dependence on
$\delta x$.\
\textcolor{R2}{ However, the
linear relationship cannot be valid in the entire range of the variation of $\delta x$.
On the one hand, the gain profile in the amplifying fiber is not as linear as the function
of the wavelength~\cite{becker1999erbium}. On the other hand, a larger variation of the wavelength
may lead to a large jump, such as in the case of the hysteresis observed in Fig. \ref{F4}(c).
Nevertheless, it is reasonable to use the linear model to fit the tunability in the small
wavelength-tuning range. Therefore, the present regime may offer an effective method to
accurately tune the wavelength of SMs.}

In the experiment, we have also found that TS of the bound solitons features
a large jump following further increase or decrease of $\delta x$. However,
the original TS is not exactly recovered when $\delta x$ returns to the
original value, i.e., hysteresis occurs. As mentioned above, it usually
takes place between power- or polarization-dependent states in mode-locked
lasers~\cite{komarov2005multistability,fedorov2019irreversible}. Our results
reveal a different phenomenon in terms of the TS in the SM states, \textit{%
viz}., the wavelength-dependent hysteresis. Figure \ref{F4}(c) shows measured
autocorrelations of SMs, with $\tilde{n}=10+19i$, as functions of $\delta x$%
. The top plot shows the situation corresponding to varying $\delta x$ from $%
-10$ to $+10$, while TS $\tau$ jumps from $4.7$ to $3$ ps around $\delta x=2$%
. The inverse jump happens around $\delta x=-2$, when $\delta x$ varies from
$+10$ to $-10$, as shown in the bottom plot of Fig. \ref{F4}(c). Further, Fig. \ref{F4}(d)
shows the TS dynamics corresponding to Fig. \ref{F4}(c), with blue circles and
magenta stars representing, respectively, the data collected in the
experiment with $\delta x$ varying from left (shorter wavelength) to right
(longer wavelength) and vice versa. Summarizing these observations, we
conclude that two different bound states with large and small values of TS
coexist in the middle green area, depending on the direction of the
variation of $\delta x$. Such bistability areas are key elements used in
optical logic devices~ \cite{smith1984optical,smith1986optical}.

\textcolor{R1}{Geneally, hysteresis is an interesting phenomenon
in mode-locked fiber laser systems. For example, the hysteresis can be
induced by variations of the power or polarization~
\cite{fedorov2019irreversible,komarov2005multistability}, and by
interactions with dispersive waves \cite{yi2017single}, while our system demonstrates
the wavelength-controlled hysteresis.} We conclude that the SM-based
hysteresis, found here in terms of the TS and controlled by the shift of the
central wavelength, may be more robust than the conventional optical
hysteresis controlled by the pulse's power, which is prone to instabilities
breaking the power balance \cite%
{herink2017real,krupa2017real,komarov2012multiple,fan2015pump}.

The wavelength tunability of the SMs and their hysteresis are promising for
the development of the optical encoding and logic setups. For instance, the
SM may be selected as $\ket{m}_{A}$ with wavelength $\lambda _{A}$, and as $%
\ket{m}_{B}$ with another wavelength $\lambda _{B}$, both having the same TS
(Fig. \ref{F1}(b)). This setting will support a larger variety of SMs in the
nonlinear dissipative system, thus expanding the available multiplicity of
the SM-based encoding and switching. Therefore, it may find applications to
optical massive data processing~\cite{feldmann2021parallel}. Further, the
possibility to control the bistability by precise adjustment of position $%
\delta x$ of the hologram in the free space may be appropriate for the
design of optical free-space computational processing~\cite{zhou2021large}.

\section{Discussion}

The objective of this work is to propose an effective scheme for
manipulations of the SMs (soliton molecules) built as bound states of two
optical solitons in a nonlinear dissipative system, i.e., the mode-locked
fiber laser. The experimental results, in agreement with systematic
simulations of the laser's model, based on the nonlinear equations of the
CGLE types, are summarized as functions of the real and imaginary parts of
the effective GVD of the laser cavity.
\textcolor{R1}{The experimental findings clearly demonstrate that
the management of the dispersion loss is an efficacious tool for tailoring
bound states in the present fiber-laser system. Namely, it allows us to preselect
the TS (temporal separation) between the bound soliton in the range of ([$3.014$,$5.478$] ps,
with the resolution of $0.112$ ps). The same tool makes it possible to accurately tune the
central wavelength of the bound state, and control the hysteresis in terms of the
wavelength.}
\textcolor{R1}{The numerical simulations reported in this work also show that the dispersion
loss is an effective control parameter, which can be used to accurately select the value of the
TS. It is worthy to note
that this parameter is also an important one on other models based on
CGLEs~\cite{malomed1991bound,akhmediev1997multisoliton,sakaguchi2018stationary,song2019recent}.
An explanation is that a slight variation of the dispersion loss affects the potential
of the interaction of the two solitons,
which leads to a shift of positions
of the interacting solitons, and may trigger jumps between bound states with different
values of the TS. This phenomenology is somewhat similar to the recent prediction
of transitions between different bound states, which includes a possibility of hysteresis
created by the third-order dispersion~\cite{sakaguchi2018stationary}. }

The results reveal not only the new degree of freedom in the experiments
with nonlinear dissipative systems in photonics, but also allow encoding
data in the form of multiple stable SMs with different TSs, and SMs with
fixed TS but different colors (central wavelengths). Such multiple states
may be toggled in a controllable fashion, which is important for
applications of SMs to communications~\cite{stratmann2005experimental},
optical switching~\cite{kurtz2020resonant}, and storage~\cite%
{leo2010temporal,pang2016all}. Also, these findings offer potential for
the use of the optical SMs in high-capacity schemes for optical data
processing~\cite{feldmann2021parallel,zhou2021large}, machine learning based
on ultrafast photonics~\cite%
{genty2021machine,iegorov2016direct,wei2020harnessing,pu2020intelligent},
and operations with high dimensional temporal quantum information~\cite%
{drummond1993quantum,kues2017chip}.

The results demonstrate that the hologram-driven mode-locking is a robust
and accurate technique for the work with the optical SMs. The switching
speed of the present setup, $\sim$ 2 Hz, is not high, being limited by the
necessity to refresh the liquid-crystal SLM (spatial light modulator). The
scheme may be improved by using faster SLM or high-speed amplitude (or
phase) modulators, such as ones based on digital micromirror devices, that
secure the speed in the range of~kHz~\cite{scholes2019structured}. It is
also possible to employ the acousto-optic modulator that typically operate
in the $\sim 100$ MHz range~\cite{weiner2011ultrafast}.
\textcolor{R1}{In the latter case, it is possible to monitor
the dynamics of switching between different SMs.}

\section{Appendix}

\subsection{Simulation of the soliton molecules}

\setcounter{equation}{0} \renewcommand\theequation{A\arabic{equation}}

The possibility of the existence of soliton bound states in the mode-locked
fiber system can be predicted by the model based on the complex
Ginzburg-Landau equation (CGLE)~\cite%
{malomed1991bound,akhmediev1997multisoliton,song2019recent}, which is
written as Eq.  (\ref{E1}) in the main text. Simulations of the CGLE equation
produce results for SMs (soliton molecules) similar to the experimental
findings. To provide still better proximity of the numerical results to
the experimental ones, we performed simulations of a more detailed model,
based on CGLEs in the fiber sections of the system, the pulse-shaping function
in the free-space section, and saturation transfer function of the
carbon-nanotube-based absorber.

The setup of the mode-locked fiber laser is shown in Fig. 1 of the main
text. It includes two fiber segments (the single-mode one and the doped gain
fiber), which provide GVD and gain. Accordingly, the model includes CGLEs
and the additional ingredient, defined as follows:
\begin{equation}
{\;\frac{{\partial A}}{{\partial z}}=-i{\frac{{{{\tilde{\beta}}_{2,\mathrm{%
fiber}}}}}{{2}}}{{{\frac{\partial ^{2}}{{\partial T^{2}}}}}}A+\frac{g}{{%
2\Omega _{g}^{2}}}\frac{{{\partial ^{2}}}}{{\partial {T^{2}}}}A+\left( {%
\frac{g}{2}+i\gamma |A{|^{2}}}\right) A.}  \label{ES2}
\end{equation}%
In the section representing the pulse shaper, the underlying equations can
be averaged and added to Eq. \ref{ES2} by addressing high order dispersions ~
\cite{runge2020pure}. However, we find that the model is made more accurate
if it explicitly includes the pulse-shaping equation. Therefore, the
free-space element (the $4f$ pulse shaper~\cite{monmayrant2010newcomer}) that
connects to the complex dispersion coefficient, is modeled by:
\begin{equation}
{A(\omega ,z+{L_{4\mathrm{f}}})=A(\omega ,z)\cdot \sqrt{\delta }\exp \left( {%
i\sum\limits_{k}{\frac{{{{\tilde{\beta}}_{k,\mathrm{holo}}}{L_{\mathrm{cavity%
}}}{\omega ^{k}}}}{{k!}}}}\right) .}  \label{ES3}
\end{equation}%
Here, ${{L_{\mathrm{cavity}}}}$ is the overall length of the cavity, ${{L_{%
\mathrm{4f}}}}$ is the overall length of the $4f$ pulse shaper,
\textcolor{R2}{$\delta $ is
the internal linear transmission rate  that mainly originates from the SLM}
(spatial light modulator), optical grating, and free-space-to-fiber
coupling. The coefficient $\tilde{\beta}_{k,\mathrm{holo}}$ represents, as
said above, the additional $k$-order complex dispersion. Recall that $\tilde{%
\beta}_{2,\mathrm{fiber}}$ in Eq. (\ref{ES2}) represents the complex
dispersion of the fiber, ignoring the higher-order terms.

To realize the stable mode-locking regime, the setup includes the homemade
saturable absorber based on carbon nanotubes, cf. Refs.~\cite%
{liu2013versatile,liu2018real}. It is characterized by the transfer
function,
\begin{equation}
T_{\mathrm{ab}}=\sqrt{1-(\alpha _{\mathrm{ns}}+\alpha _{0}/(1+A^{2}/I_{%
\mathrm{sat}}\cdot A_{\mathrm{eff}}))}.  \label{ES4}
\end{equation}%
Here $\alpha _{\mathrm{ns}}$ is the unsaturated absorption, $\alpha _{0}$ is
the linear limit of the saturable absorption, $I_{\mathrm{sat}}$ is the
saturation intensity, and the effective fiber's cross-section area is $A_{%
\mathrm{eff}}=63.6\times 10^{-12}~$m$^{2}$.%

Using Eqs. (\ref{ES2})-(\ref{ES4}), we simulated the evolution of the
soliton in the cavity. In particular, CGLE given by Eq. (\ref{ES2}) was used
to simulate the propagation through the fiber; Eq. (\ref{ES3}), with the
appropriate GVD coefficient, described the action of the $4f$ pulse shaper
in the frequency domain; and the transfer function defined as per Eq. (\ref%
{ES4}) determined the passage of the pulse through the saturation absorber, $%
A\rightarrow A\cdot T_{\mathrm{ab}}$. Coefficient $g$ in Eq. (\ref{ES2}) is
the gain parameter for the erbium-doped fiber, which is determined by the
small-signal gain $g_{0}$ and saturation energy $E_{\mathrm{sat}}$:
\begin{equation}
g=g_{0}\exp (-E(z)/E_{\mathrm{sat}}),  \label{ES5}
\end{equation}%
where $E(z)=\int_{-\infty }^{+\infty }|A(z,t)|^{2}dt$ is the total energy of
the pulse traveling in the cavity.

The element of the setup described by Eq. (\ref{ES3}) performed pulse
shaping in the time domain by means of the hologram encoding the
frequency-domain features. The optical grating transforms the input from the
time domain into its counterpart in the frequency domain. Then, the convex
lens focused different frequencies at different spatial locations, thus
introducing the spatial dispersion in the Fourier plane. With the help of
the conjugate system, the pulse was then transformed from the spatial domain
back into the temporal one. The central frequency of the $4f$ pulse shaper,
along with the surface of the SLM, follows this simple relationship based on
the diffraction equations ~\cite{weiner2011ultrafast,monmayrant2010newcomer}:
\begin{equation}
\frac{{\partial \lambda }}{{\partial x}}=\frac{{d_{\mathrm{grating}}\cos ({%
\theta _{\mathrm{diff}}})}}{{mf}}  \label{ES6}
\end{equation}%
Here, $m$ is the diffraction order (usually $m=1$), $\theta _{\mathrm{diff}}$
is the input angle of the optical grating (here, it is $47.5^{\circ }$), $%
f=125~$mm is the focal length of the input lens, and $d_{\mathrm{grating}}$
is the grating constant, which is $940$ per mm in our experiment. Following
these definitions, the corresponding spatial dispersion of the SLM is
expected to be $5.75$ nm$/$mm, while, in our experiment, the measured
spatial dispersion is $5.2$ nm$/$mm.
\textcolor{R2}{Specifically, we set a double-slit
hologram in the SLM to calibrate the actual
spatial-dispersion relationship $\partial \lambda / \partial x$.  The spatial separation between two slits corresponds to the
frequency (wavelength) distance between two peaks from the output spectrum.}
In the simulations, we also needed to consider the filtering effect of the
SLM's surface size, where we set the spectrum bandwidth to be $45$ nm. Due
to the consideration of this term, we ignored the gain bandwidth of $\Omega
_{g}$ in Eq. (\ref{ES2}). \setcounter{figure}{0} \renewcommand\thefigure{S%
\arabic{figure}}
\begin{figure*}[t]
\centering
\includegraphics[width=16cm]{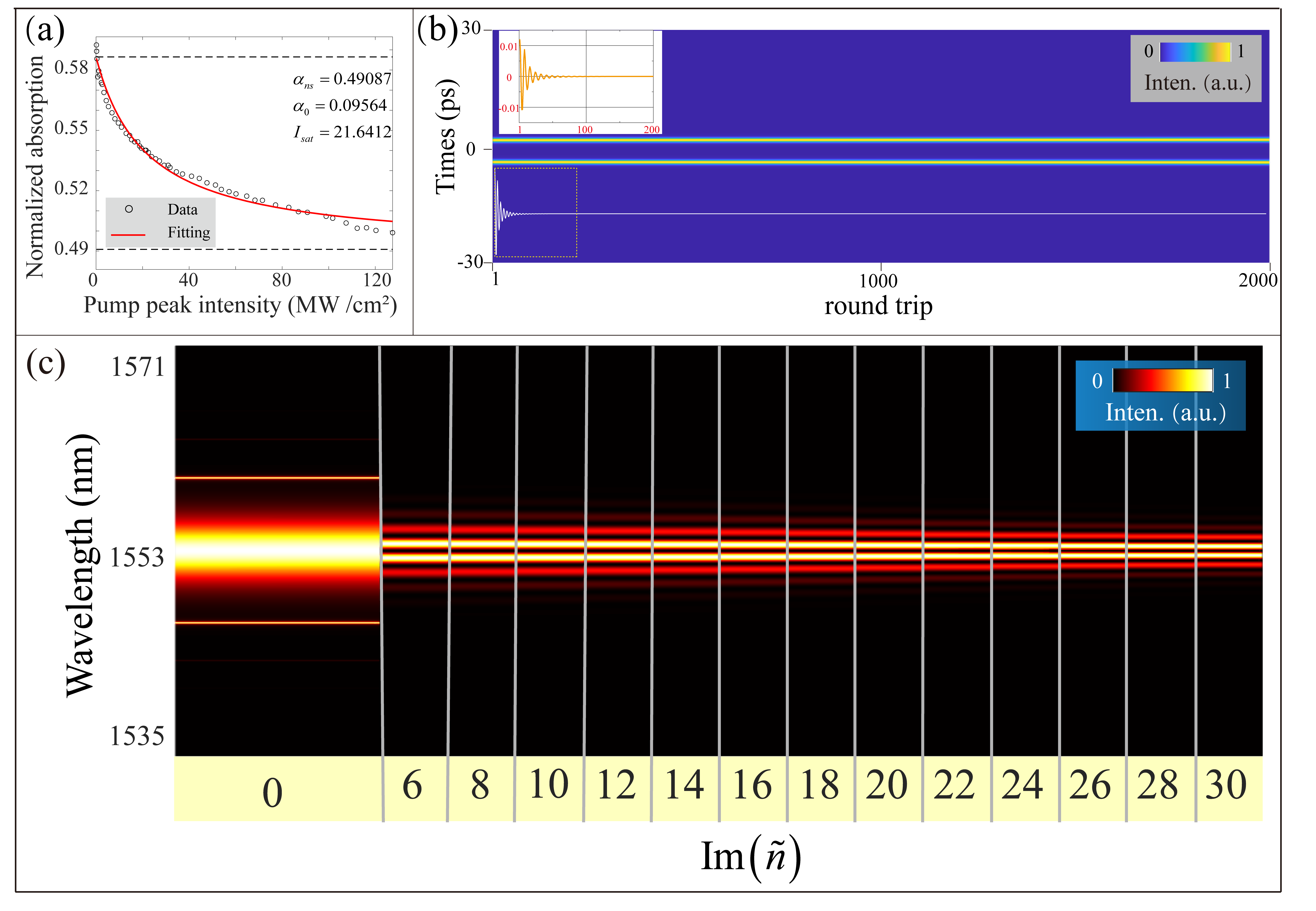}
\caption{The absorption curve for the carbon nanotube saturable absorber (a)
and evolution of the output pulse (b) produced by the fiber cavity. In panel
(a), the fitting parameters are $\protect\alpha _{0}=0.095$, $\protect\alpha %
_{\mathrm{ns}}=0.4908$, $I_{\mathrm{sat}}=21.6412$ W/m$^{2}$. In panel (b),
the SM's output intensity is displayed, for $\tilde{n}=10(1+i)$; the
horizontal axis represents the number of round trips from $1$ to $2000$; the
vertical coordinate shows the temporal coordinate from $-30$ to $+30$ ps.
Here, the separation between the bound solitons takes a stable value, $%
\protect\tau =5.67$ ps. \textcolor{R2}{The bottom portion of the panel (b)
shows
the difference in the pulse energy
between the current and last
round
trips. The top inset shows details for the evolution
between $1$ and
$200$
round trips, which corresponds to the plot shown in the bottom portion.}
Panel (c) shows all normalized spectra intensities of the final outputs,
where the horizontal axis represents the sequence of values of $\mathrm{Im}(%
\tilde{n})$, and the vertical axis is the wavelength, the central one being $%
1553$ nm. }
\label{FS1}
\end{figure*}

The parameters used in the simulations were chosen as close as possible to
their experimental counterparts. The length of the erbium-doped fiber is
$3.3$ m, with $\mathrm{Re}(\tilde{\beta}_{2,\mathrm{gain-fiber}})=20$ ps$%
^{2}/$km and $\gamma =0.0032$, the length of the single-mode fiber is $15.48$
m, with $\mathrm{Re}\left( \tilde{\beta}_{2,\mathrm{single-mode-fiber}%
}\right) =-21.6$ ps$^{2}/$km and $\gamma =0.0013$, and the net dispersion of
the entire fiber is $-0.2672~$ps$^{2}$. Here, as said above, we ignore the
higher-order complex dispersion in the fiber section. The length of the
free-space path is $1.1$ m, hence the total length of the cavity is $19.88$
m. In our setup, the single-mode fiber is placed in front of the $4f$ pulse
shaper, therefore the simulations were performed in two main parts of the
setup, \textit{viz}., the fiber and free space. For the erbium-doped fiber,
the small-signal gain is $2.75$, and $E_{\mathrm{sat}}=63$ pJ for the
single-soliton state (the value is fitted by matching the bandwidth); for
the SMs, $E_{\mathrm{sat}}=80~$pJ is set (this value is fitted by matching
the TS in the SMs); and the internal linear transmission $\delta $ is set to
be $0.10$. For the homemade carbon-nanotube saturable absorber, we measured
its absorption curve, which is shown in Fig. \ref{FS1}(a). By fitting the
data with the help of the absorber equation, we obtained the following
parameters: $\alpha _{0}=0.095$, $\alpha _{\mathrm{ns}}=0.4908$, and $I_{%
\mathrm{sat}}=21.6412$ MW/cm$^{2}$.

For the $4f$ pulse shaper, which is modeled by Eq. (\ref{ES3}), we set $%
\mathrm{Re}(\tilde{n})=10$, and vary $\mathrm{Im}(\tilde{n})$ from $6$ to $%
30 $, which determines $\tilde{\beta}_{k}$, as defined above. Because there
maybe a small discrepancy between the actual and theoretical parameters of
the holograms, we directly use the hologram's phase to add the complex
dispersion, which may include higher-order terms (See \textbf{Section 2 in
Supplementary Material} for the discussion of the small discrepancy).

The production rate of the output coupler of the fiber cavity is set to be $%
10\%$, and the overall internal loss of the $4f$ pulse shaper is about $90\%$%
, due to the use the phase-amplitude encoding technique \cite%
{bolduc2013exact,liu2019classical}. With these parameters, we have performed
the simulations of the system for $2000$ round trips. In the simulations, we
first generated a conventional soliton $A_{\mathrm{soliton}}(T)$ from the
random noise $\mathrm{Rand(N)}$ that which is followed by the Gaussian
distribution [$=\mathrm{Rand(N)}\times \mathrm{\exp }(-(T/2)^{2})$], keeping
the bandwidth of the simulations as close as possible to the experimental
value. Then, we used the individual pulses to build two-soliton pairs [${A_{%
\mathrm{soliton}}}(T-\tau _{0}/2)+{A_{\mathrm{soliton}}}(T+\tau _{0}/2)$, $%
\tau _{0}$ being their relative separation] as the input for the next stage
of the simulations, with $\tilde{n}=10+6i$; in this case, the output
produces a stable bound state. This output is then used as the input for the
next iteration. Fig. \ref{FS1}(b) shows the output fields for $\tilde{n}%
=10\left( 1+i\right) $, between $1$ and $2000$ round trips.
\textcolor{R2}{For testing the stability, we calculated the difference in
the pulse energy between the given round trip and the last one, which is
displayed in the bottom portion of Fig. \ref{FS1}(b). The top inset in the same figure
shows the results for the round trips from $1$ to $200$. Monitoring the difference in the
pulse energy, we have concluded that the bound state remains stable after
an initial stage of strong oscillation.} We extract the last output field
after $2000$ round trips and calculate their spectral intensity. All
normalized spectral intensities of the final outputs are separately shown in
Fig.~\ref{FS1}(c). We conclude that the spectral fringes decrease with the
increase of the imaginary part of the GVD coefficient, $\mathrm{Im}(\tilde{n}%
)$, which implies that TS of the SMs grows accordingly. The Fourier transform
of these final output spectra produces the autocorrelation function which is
displayed in Fig. 2(g) of the main text.

The simulations were performed to verify the physical mechanism. The results
produced by the current theoretical model do not agree with experiments
exactly, as some simplifications are used in it.
\textcolor{R2}{ For example, the difference in separations between the
experiments and simulations are due to a minor difference between them
in the GVD and dispersion loss. The difference in the observed relative phase
of the bound solitons is chiefly affected by the pump power and
polarization~\cite{herink2017real,gui2018soliton}.} In addition,
\textcolor{R2}{in the simulation, we mainly focused on the effect of the GVD
and frequency-dependent loss on the TS in the bound states, and we did not
consider in detail the effects of the polarization. Actually, the SLM can
operate only in horizontal polarization. Thus, in the stable dynamical regime,
only the horizontal polarization is effectively modulated, while the vertical polarization
component is strongly attenuated by losses.} Nevertheless, the main features
observed in the experiments are correctly predicted by the present model, in
which the values of the TS in the SMs increase with the growth of the
imaginary part of the GVD, $\sim\mathrm{Im}(n)$.

\subsection{The pulse shaper system and hologram}

For the generation of a target arbitrary electric component of the optical
field, $E_{\mathrm{out}}(x,y)=A(x,y)\exp (i\phi )$, with a certain amplitude
and phase, the hologram needs to include the information concerning the
amplitude and phase. The hologram could be synthesized by the
amplitude-phase encoding, defined as \cite%
{davis1999encoding,bolduc2013exact,liu2019classical}:
\begin{equation}
\Phi (x,y)_{\mathrm{holo}}=\mathrm{sinc}(\pi -M\pi )\cdot \mathrm{Arg}%
(E)\cdot \Lambda (x,y),  \label{ES7}
\end{equation}%
where $M$ $(M=|A(x,y)|,0<M<1$) is a normalized constraint for the positive
output amplitude function, and $\mathrm{Arg}(E)=\exp (i(\phi ))$ is the
phase of the output complex field. Further, $\Lambda (x,y)$ is an optical
blazed grating phase, that may be $\exp \left[ {2\pi i\left( {{G_{x}}x+{G_{y}%
}y}\right) }\right] $, where $G_{x,y}$ are spatial frequencies along the $x$
and $y$ directions. In our experiment, we used only the optical grating
oriented along $y$. The grating is necessary to avoid the interference with
effects of the zero-order diffraction in the blazed grating phase.

To realize the modulation of the amplitude and phase in our study, i.e., $%
H(x)=\phi _{R}(\mathrm{Re}(\tilde{\beta}_{2,\mathrm{holo}}),x)\cdot \phi
_{I}(\mathrm{Im}(\tilde{\beta}_{2,\mathrm{holo}}),x)\cdot \phi _{G}(d,x)$,
the amplitude (determined by $M$ in Eq. \ref{ES7}) should be taken as $\phi
_{I}(\mathrm{Im}(\tilde{\beta}_{2,\mathrm{holo}}),x)\sim \ \exp \left( {-{%
k_{I}\mathrm{Im}(\tilde{n}){L_{\mathrm{cavity}}}\omega ^{2}}/2}\right) $,
the phase information ($\mathrm{Arg}(E)$ in Eq. \ref{ES7}) is given by $\phi
_{R}(\mathrm{Re}(\tilde{\beta}_{2,\mathrm{holo}}),x)\sim \exp \left[ {i{k_{R}%
\mathrm{Re}(\tilde{n})}{L_{\mathrm{cavity}}}{\omega ^{2}}/2}\right] $, and
the optical blazing ($\Lambda (x,y)$) is $\phi _{G}(d,x)=\exp (-i2\pi Gy)$.
Here, $k_{R,I}$ are constant coefficients of the real and imaginary parts,
respectively. In actual encoding, we replace frequency $\omega $ by spatial
coordinate $x$, using the dispersion relationship of the $4f$ pulse shaper,
see Eq. (\ref{ES6}). Note that the current encoding format leads to a small
difference in the beam's profile in comparison to the target profile \cite%
{bolduc2013exact}, due to higher-order effects. Details of the $4f$ pulse
shaper, the hologram structure and high-order effects are presented in
\textbf{Section 1 and 2 in Supplementary Material}.

\begin{backmatter}
\bmsection{Funding} This work was supported by the International Postdoctoral Exchange Fellowship Program 2020 by the Office of China Postdoctoral Council (No. 34 Document of OCPC,2020), the National Natural Science Foundation of China under Grant Agreements 61705193, Fundamental Research Funds for the Central Universities 2019QNA5003, Natural Science Foundation of Zhejiang Province under Grant Nos. LGG20F050002.
And in part, by the Israel Science Foundation through grant No. 1286/17. Also, Shilong Liu and Ebrahim Karimi acknowledge the support of Canada Research Chairs (CRC) and Canada First Research Excellence Fund (CFREF) Program.

\bmsection{Acknowledgments}
We thank Prof. Xueming Liu (Zhejiang University) for providing the part of the experimental platform, Prof. Baosen Shi and Dr. Zhiyuan Zhou (University of Science and Technology of China) for providing the infrared spatial light modulator and BBO crystal, Lin Huang (Zhejiang University)   for discussion of the results, and Ruimin Jie also Mengqin Gao (Zhejiang University) for helping to measure parts of data in experiments. Also, appreciation for Ende Zuo (University of Ottawa) for improvement of figure esthetics.

\bmsection{Disclosures} The auhors declare no conflicts of interest.

\bmsection{Data availability} Data underlying the results presented in this paper are not publicly available at this time but may be obtained from the authors upon reasonable request.

\bmsection{Supplemental document}
See Supplement 1 for supporting content.

\end{backmatter}

\section{References}

\end{document}